\documentclass[prd,twocolumn,superscriptaddress,altaffilletter,lengthcheck, tightenlines]{revtex4}
\usepackage{graphicx}
\usepackage{bm}
\newcommand{\be}{\begin{equation}}
\newcommand{\ee}{\end{equation}}
\newcommand{\ben}{\begin{eqnarray}}
\newcommand{\een}{\end{eqnarray}}
\newcommand{\n}{\label}
\newcommand{\no}{\noindent}

\newcommand{\ke}{$k$-essence }
\newcommand{\kf}{$k$-field }

\newcommand{\ga}{\gamma}
\newcommand{\ep}{\epsilon}
\begin{document}

\author{L.P. Chimento}
\email{wtpagagj@lg.ehu.es}
\affiliation{Dpto. de F\'\i sica, Facultad
de Ciencias Exactas y Naturales, Universidad de Buenos Aires,
Ciudad Universitaria, Pabell\'on I, 1428 Buenos Aires,
Argentina}
\author{M\'onica Forte}
\email{monicaforte@fibertel.com.ar}
\affiliation{Dpto. de F\'\i sica, Facultad
de Ciencias Exactas y Naturales, Universidad de Buenos Aires,
Ciudad Universitaria, Pabell\'on I, 1428 Buenos Aires,
Argentina}
\author{Ruth Lazkoz}
\email{wtplasar@lg.ehu.es}
\affiliation{Fisika Teorikoa, Zientzia eta Teknologiaren Fakultatea, Euskal Herriko Unibertsitatea, 644 Posta Kutxatila, 48080 Bilbao}
\title{Dark matter to dark energy transition in k-essence cosmologies}


\begin{abstract}

We implement the transition from dark matter to dark energy in 
k-essence cosmologies for a very large set of kinetic functions $F$, in a
way alternative to recent proposals which use generalized Chaplygin gas and
transient models. Here we require that the pressure admits a power-law
expansion around some value of the kinetic energy where the pressure vanishes.
In addition, for suitable values of the parameters of the model, the speed of
sound of the dark matter  will be low. We first present the discussion in
fairly general terms, and later consider for illustration two  examples.

\end{abstract}
\maketitle
\section{Introduction}

According to everyday gathered more and better astrophysical evidence ,
 the evolution of the Universe  is largely driven by dark energy  with negative
 pressure together with pressureless cold dark matter (see \cite{sahni} for the latest review). However little is know about the origin of either component, which in the standard cosmological model would
 play very different roles: dark matter would be responsible for matter clustering, whereas dark energy would account for accelerated expansion. This lack of information
leaves room to speculate with the idea that a single component acted in fact as both dark
 matter and dark energy.  
 
 On one hand the unification of those two components makes model building becomes  considerable simpler, but on the other hand, and more importantly, it implies the existence of an era during which the energy densities of dark matter and dark energy are  strikingly similar. One should therefore be not surprised that unifying dark matter energy has risen a considerable theoretical interest. 

The first two unifying candidates to appear in the literature were 
the Chaplygin fluid and the tachyon field but here we will consider an emerging alternative: k essence.
Specifically, in this paper we revisit the idea put forward in \cite{scherrer} that scalar field models with non-canonical standard kinetic terms can provide a unified description of dark matter and 
dark energy. Given that the equations of motion in  classical theories seem to be of second order, the only kinetic terms considered will be functions of the square of the gradient of the scalar field (hereafter k field). Moreover, since k fields can be used for devising dark energy models, it   is common place to interpret those fields as some kind of matter called k essence   \cite{scherrer,k-essence,tracking}. Nevertheless, k fields were not originally introduced for describing late time acceleration, but rather they were put forward  as possible  inflation driving agents \cite{kinflation,kinfper}. Interestingly enough, as shown in \cite{tracking}, one can also  construct tracking k essence cosmologies.

In \cite{scherrer} it was shown that, under some assumptions on the form of the Lagrangian
of k essence models it is possible to obtain examples of universes which transit between
a dark matter dominated era and a dark energy dominated one. In that paper, the author considered Lagrangians
with a constant k-field potential (i.e., Lagrangians which functionally depend only on field derivatives and not on the field itself), and it was argued that if the pressure had an extremum at a given value of the field derivative, then the energy density of the model would scale like the sum of a non-relativistic dust component and a cosmological constant-like component. 

By sticking also with Lagrangians with a constant k-field potential,  we obtain the new and
interesting result that the transition between dark matter and dark energy can also be successfully modeled  without having to require  the presence of a extremum in the pressure. Moreover, the parameters of the model can be chosen so that the sound speed is very low,
thus letting the pressureless component behave as dark matter. We illustrate our results
with two interesting examples: the first one is a quadratic toy model with necessarily large  speed of sound, but the second one is a transient model 
in which the speed of sound may be as small as one wishes.

\section{Basic equations} 

Since we are interested in a large-scale description of the universe, we
assume k essence is the source of a spatially flat homogeneous and isotropic
spacetime with scale factor $a(t)$ and Hubble factor $H=\dot a/a$ (as usual, here and throughout overdots denote differentiation with respect to $t$). Our models are derived from the
factorizable Lagrangian $
{\cal L}=-V(\phi) F(x)$,
where $V(\phi)$ is a positive definite potential, $F(x)$ is an arbitrary
function of $x$, $\phi$ is the k field, and $x=-\dot\phi^2$.
That form of the Lagrangian  is in turn suggested by the Born-Infeld  Lagrangian
${\cal L}=-V(\phi)\sqrt{1+x}$
which was associated with the tachyon  by computations in boundary string
field theory \cite{BSFT}. Such Lagrangian also arises in open bosonic string theory \cite{fratse} and is a key ingredient in the effective
theory of D-branes \cite{Leigh}.

Associating the energy-momentum tensor of the k field 
with that of a perfect fluid, we compute  the energy
density $\rho$ and the pressure $p$, which read
\ben
\n{ro}
&&\rho=V(\phi)[F-2xF_x],   \\  &&\n{p} p=-V(\phi)F.
\een
The corresponding barotropic index $\gamma$ is given by
\ben
\n{ga}
&&\gamma\equiv1+p/\rho=-{2xF_x}({F-2xF_x})^{-1}.
\een
On the other hand, the Einstein field equations are 
\ben
\n{00}
&&3H^2=V[F-2xF_x],\\
\n{11}
&&\dot H=VxF_x,
\een
\no whereas the conservation equation reads
\be
\n{con}
\dot\rho+3H(\rho+p)=0.
\ee
 Inserting  Eqs. (\ref{ro}) and (\ref{ga})
into the conservation equation (\ref{con}), we find the field equation for the
\kf:
\be
\n{kg}
\big({\gamma}/{\dot\phi}\big)\dot{\,}+
3H(1-\gamma)\big({\gamma}/{\dot\phi}\big)+
{V'}(1-\gamma)/{V}=0.
\ee
The stability of k essence with respect to small wavelength perturbations
requires that the effective sound speed \cite{kinfper} 
\be
\n{c}
c_s^2=
{p_x}/{\rho_x}=
{F_x}({F_x+2xF_{xx}})^{-1},
\ee
\no be positive. However, in \cite{car} it was shown that a positive
sound speed is not a sufficient condition for the theory to be stable. 
This is all so far in what the preliminaries are concerned. In the next section
we turn our attention to the description of unifying dark matter energy
using purely kinetic k essence.

\section{K essence with a constant potential}

The first ever studied k essence models had a constant potential; the inflationary
behaviour they described had, then, a purely kinetic origin and it was dubbed k inflation \cite{kinflation,scherrer,Chimento}. Specifically, in such cosmologies inflation is pole-like, that is, the scale factor evolves like a negative power of time.
An earlier theoretical framework in which
(pole-like) k acceleration arises naturally is the pre-big bang model of  string cosmology \cite{prebigbang}. In this setup, acceleration  is just due to a scalar field called the dilaton, and it will only manifest itself in the string conformal frame. Finally, for other ideas on kinetic inflation one may have a look at \cite{levin}, where acceleration was put down to a dynamical Planck mass. 

Coming back to k essence, for a constant potential $V=V_0$, the k field equation (\ref{kg}) admits the first
integral
\be
\n{pi}
a^3F_x\dot\phi=
{6c}/{V_0}\label{first_int},
\ee
\no where $c$ is an arbitrary integration constant. Alternatively, Eq. (\ref{first_int}) can be written as
\be
\n{h.}
\dot H=-
{6c\dot\phi}/{a^3},
\ee
after using Eq. (\ref{11}). Combining Eq. (\ref{first_int}) with
the Friedmann equation (\ref{00}), the barotropic index associated with this
kind of \ke can be written in a more convenient form
\be
\n{gam}
\gamma=({1+V_0^2a^6FF_x/72c^2})^{-1}.
\ee
\no We anticipate that  there is a large set of models which describe 
universes which are dust dominated in their early stages, that is, $\gamma\approx 1$ or equivalently $p\approx 0$, which according to
the conservation equation  (\ref{con}) implies that $\rho\approx
a^{-3}$. These models are precisely generated by the set of functions $F$
which at early times satisfy the condition $a^6FF_x\ll 1$. By construction,
such models will behave at intermediate times  as if they were filled with a
perfect fluid with equation of state $p\propto\rho$. Finally, such universes
end in a stable de Sitter accelerated expansion scenario. In a way, these
models play the same role that the generalized Chaplygin gas, i.e., they
interpolate between dark matter at early times and dark energy at late times.
However, they do not share some of the unwanted features of Chaplygin
cosmologies. We dwell now on the details of the construction of those
cosmologies.

In a recent paper \cite{scherrer}  a class of models with functions $F(x)$
such that they have an extremum were considered; specifically they obey 
\be
\n{fs}
F(x)=F_0+F_2(x-x_0)^2,
\ee
\no where $F_0$ y $F_2$ are constants. The extremum is located at $x_0$ and if
$\vert x-x_0\vert\ll 1$  then $F_x\approx 0$.  Moreover, because of Eq.
(\ref{gam}) we will have $\ga\approx 1$ and the model will behave at early
times as if its matter content were dust (pressureless matter).

The main goal of this paper is to show that the hypothesis that the model is
generated by an $F(x)$ with an extremum is very restrictive. Indeed, if we
take the class of functions which admit a  expansion in powers in the form
\be
\n{ft}
F(x)=F_0+ F_1(x-x_0)+F_2(x-x_0)^2+....,
\ee

\no and then look at the definition of the barotropic index (\ref{gam}), we
see that near $x_0$ the condition $a^6FF_x\approx 0$ leads to $\ga\approx 1$,
obtaining a matter filled universe. Thus, there are two options for getting a
dust cosmology: either making $F(x_0)=0$, that is, taking $F_0=0$, where
$x_0=x(t_0)$, or making $F_x(x_0)=0$ as suggested by Scherrer, that is, taking
$F_1=0$. To investigate the first option we define the small quantity $\ep$ as
\be
\n{ep}
x=x_0(1+\ep).
\ee
\no Then, using equations (\ref{00}), (\ref{ft}) and (\ref{ep}) we can compute
the energy density up to the first order in   $\ep$. We get
\be
\n{ro1}
\rho\approx V_0x_0\left[-2F_1-(F_1+4x_0F_2)\,\ep+...\right].
\ee
\no On the other hand, inserting Eqs. (\ref{ft}) and  (\ref{ep}) into the first
integral (\ref{pi}) we find an expression for $a^{-3}$ which depends on $\epsilon$
and such that combined with Eq. (\ref{ro1}) results in
\be
\n{rof}
\rho\approx 
{12c\sqrt{-x_0}}/{(V_0\,a^3)}.
\ee

\no Remarkably, we get a dust-like evolution, without making any assumption on
$F_1$ and $F_2$. Nevertheless, at zeroth order in $\ep$ the sound speed reads
$c_s^2\approx F_1/(F_1+4x_0F_2)$ and its value does indeed depend on $F_1$ y
$F_2$. If we now impose  $4x_0F_2\gg F_1$ we get $c_s^2\approx 0$, that is, we
can describe matter which is easily concentrated by gravity.
In addition, from Eqs. (\ref{00}) y (\ref{11}) we get
\ben
\n{F'}
&&F_x(x_0)=-{3H^2(x_0)}/{2x_0V_0}\\
\n{.H}
&&\dot H(x_0)=-{3}/{2}H^2(x_0).
\een

\no If we compare now with an evolution $a=t^{2/3}$, for which $H=2/3t$ and
$\dot H=-3H^2/2$, at $x=x_0$ it follows that $t_0^2=-2/3x_0V_0F_1$. For our choice $F(x_0)=0$ the function $F$ leads to an energy density
$\rho\propto a^{-3}$ and a scale factor that matches $a=t^{2/3}$ up to the
second derivative at a $t_0$, which may be chosen arbitrarily by choosing the
value of $F_1$. However, for the option $F_x(x_0)=0$, we find
$F_0={3H^2}/{V_0}$, $\dot H(x_0)=0$ and $t_0^2={4}/{3V_0F_0}$ showing that the
main difference with respect to our option is that in this case the scale
factor only matches $a=t^2/3$ at $x_0$ up to the first derivative. Therefore,
it is the first option that best describes matter.


In the case of a constant potential and for an expanding universe $H>0$, the
variable $\Gamma=\gamma/\dot\phi$ of Eq. (\ref{kg}) is a Liapunov function for
any function $F$ provided that $\dot\phi>0$ is a positive function (if
$\dot\phi<0$, then we must take the variable $-\Gamma$). This means that $\gamma$
is a decreasing positive function, so the particular solution $\Gamma=0$,
which corresponds to the de Sitter evolution is stable whenever the
barotropic index is restricted to $0\le\gamma<1$. Here, we have improved the
results obtained in \cite{scherrer} where  the stability of de
Sitter solutions was  only proved for functions F which have an extremum. 

Our model behaves as
a sum of dark matter Eq. (\ref{rof}) with equation of state $p=0$ (see Eq. (\ref{rof}) and a
cosmological-constant-like component $p=-\rho=V_0(F(x_s)-2x_sF_x(x_s))$, where
$x_s=\lim_{t\to\infty }x$. From 
$t=t_0$ onwards (where $x(t_0)=x_0$), the matter content of the
model will mimic successively all possible $p=(\gamma-1)\rho$ fluids between
$p=0$ (dust) and $p=-\rho$ (dark energy). So, the model interpolates between
these two phases. For $t<t_0$, the energy density of the k-essence fluid behaves as
$\rho\approx a^{-3\gamma}$ with $\gamma>1$.

\section{An exactly solvable toy model}

The main results we just obtained can be illustrated with an exactly solvable quadratic model.  To that end we put forward
the  quadratic function $F(x)$ 
\be
\n{f2}
F=\frac b6+x-\frac{x^2}{2b},
\ee
which becomes null at $x=b(1\pm {2}/\sqrt{3})$
and has a extremum at $x=b$, where $b$ is a free parameter of the model. Inserting 
Eq. (\ref{f2}) into Eq. (\ref{00}) we arrive at
\be
\n{.p}
\dot\phi^2=-
{b}/{3}\pm\sqrt{
{2b}/{V_0}}\,H.
\ee
\no We then substitute the latter into Eq. (\ref{h.}) and by integration we get the relative expansion:
\be
\n{h}
H=\pm\left[\sqrt{{bV_0}/{18}}+
{9c^2}\sqrt{
{2b}/{V_0}}\,\eta^2\right],
\ee
\no where the a new parameter $\eta=\int{dt/a^3}$ has been introduced (in addition, an arbitrary integration constant in the definition of $\eta$ has been adjusted
so that the expression scale factor coincides asymptotically at large $t$  time with the one corresponding to a de Sitter solution). Now, combining the last two equations we get
\be
\dot\phi^2=
{18c^2b}\eta^2/{V_0},\label{sfeta}\ee
and the scale factor can be obtained integrating  Eq. (\ref{h}) to yield
\be
\n{af}
a^3=-\left({\sqrt{
{bV_0}/{2}}\,\eta+
{9c^2}\sqrt{
{2b}/{V_0}}\,\eta^3}\right)^{-1},   \quad \eta\le 0,
\ee
\no where the singularity has been set at $\eta=-\infty$.
The equation of state and the sound speed of this toy model are respectively
\ben
\n{ptoy}
&&p=-
{4bV_0}/{9}+
{\rho}/{3}+
\sqrt{
{8b}/{27V_0}}\,\,\rho^{1/2},\\
\n{ctoy}
&&c_s^2=
{dp}/{d\rho}=
{1}/{3}+
\sqrt{
{2bV_0}/{27}}\,\,\rho^{-1/2}.
\een
Now, by making a time shift in $\eta$ so that $\eta=\delta-\delta_0$, it will possible to cast 
(\ref{af}) in the form
\be
\n{aa}
a^3=-\left({a_0+a_1\delta+a_2\delta^2+a_3\delta^3}\right)^{-1},
\ee
\no where $a_0$, $a_1$, $a_2$ y $a_3$ are constants. If we  get closer to the singularity (i.e., $\delta\to\infty$) then 
$a^3\propto\delta^{-3}$, and taking into account that  $d\eta=d\delta$ it turns out that
$t\propto\delta^{-2}$ and the universe starts off like $a\propto t^{1/2}$. If we now move
away from the singularity and approach
$x_0$ then
$a^3\propto\delta^{-2}$, which implies $t\propto\delta^{-1}$, and the scale  factor satisfies $a\propto t^{2/3}$, as corresponds to a matter field universe.
Finally, if we keep on moving away from the singularity to reach the last epoch of the universe then $a^3\propto\delta^{-1}$ and $t\propto
-\ln{\delta}$, and  the scale factor satisfies
$a\propto\exp{\sqrt{bV_0/18}\,t}$, as corresponds to a de Sitter model. 

The equation of state tells us that initially we have a radiative fluid
$\rho\approx \rho/3$ (at high energies), but as the model proceeds to the asymptotic regime the pressure tends to a constant value and $p=-\rho=-bV_0/6$, so the fluid acts like a cosmological constant. The fluid interpolates between these two limits and passes through a dust-dominated epoch when $p$ becomes null; by the definition of $p$ as a function of $F$ this will happen at
$x=x_0=b(1-2/\sqrt{3})$ for
$b>0$ (recall that by definition $x<0$). In what concerns the speed of sound, we know it initially takes
the value corresponding to radiation, i.e. $c_s^2=1/3$. From there on it grows monotonically till it reaches the upper limit
$c_s^2=1$ at the last stage in which the evolution is de Sitter-like ($H=\rm{cons}$) and the energy density takes the limiting value $\rho= bV_0/6$. Another way to deduce this result 
is to note from combining Eqs. (\ref{sfeta}) and (\ref{af}) that on the $\eta\to 0$ limit 
 we have $x\propto a^{-6}$, and the Einstein equation
 (\ref{00}) comes down to the one corresponding to a free scalar field, that is, our k essence behaves
 like a stiff fluid with $c_s^2\approx 1$.
 
 Now, since the speed of sound is large at the stage where the evolution 
 is dust-like, there would either be a blow-up of the perturbations or excessively damped oscillations \cite{sandvik}. Nevertheless, even though this models has not much physical merit,  it is mathematically interesting because it shows the possibility of having  k essence which behaves like dust when the pressure is not extremal. 
 
\section{A transient model}

Now, we investigate a simple transient kinetic \ke model introduced in
\cite{Chimento}. It is generated by the following function $F$
\be
\n{1}
F=\beta \,V_0^{-1}\left[2\alpha \alpha_0\sqrt{-x}-(-x)^\alpha\right],
\ee
\no where $\alpha$ and $\alpha_0$ are two real constants and $\beta=1/(2\alpha-1)$. The energy density
and the equation of state of the \kf are calculated from Eqs. (\ref{ro}) and
(\ref{1})
\be
\n{1f}
\rho=(-x)^\alpha,   \qquad p=\beta\left[\rho-
{2\alpha \alpha_0}{\rho^{1/2\alpha}}\right],
\ee
\no and the sound speed becomes
\be
\n{cs}
c_s^2=\beta\left[1-
{\alpha_0}{\rho^{-1/2\alpha\beta}}\right].
\ee
\no Solving the conservation equation (\ref{con}), we obtain the energy
density in terms of a
\be
\n{rom}
\rho=\left[\alpha_0+{c_0}/{a^3}\right]^{2\alpha\beta},
\ee
\no where $c_0$ is a redefinition of the integration constant $c$. Now we will apply the
results obtained in the last section and we will expand the function $F$ in powers
around $x_0=(2\alpha\alpha_0)^{2\beta}$ where $F$ and $p$ become null. 
In this case we can evaluate  $\rho(x_0)$ using  Eq. (\ref{1f}) and then calculate the 
speed of sound given by  Eq.
(\ref{cs}). The result is
$
\n{cf}
c_s^2={1}/{2\alpha}$
\no whereas for the barotropic index we get $\gamma(x_0)=1$. Thus, for large
$\alpha$, the model is cold dark matter dominated at $x=x_0$, with
approximated vanishing sound speed, $c_s^2\approx 0$, while from Eq.
(\ref{rom}) the energy density of the transient k-essence fluid becomes $\rho\approx
\alpha_0+c_0/a^3$. In addition, at late times the model ends in a de Sitter
stage. Such ``transient model'' may be considered as an alternative model to the
generalized Chaplygin gas. It allows the evolution of the initial
perturbations in the energy density into a nonlinear regime and near $x_0$ it could play the
role of cold dark matter  (i.e. dark matter unable to resist gravitational clumping). Finally, the model yields an energy
density which scales like the sum of a non-relativistic dust component at
$x_0$ with equation of state $p = 0$ and a cosmological-constant-like
component $p=-\rho$.

\section{Conclusions}

The Universe at present seems to be expanding because of the joint action of
dark energy  with negative pressure and pressureless cold dark matter.
However, it is uncertain where this dark energy could come from, and it
is therefore fair to speculate  with the possibility of modeling those two
components by a single one: k essence. Following that line we
present a new way of using k essence for modeling  a transition form dark
matter to dark energy.

A earlier discussed by Scherrer \cite{scherrer},   that transition can be
implemented using models such that their pressure has a extremum, but our
results prove that such restrictive assumption is in fact not necessary.
Specifically, it suffices that the  pressure (as a function of the kinetic
energy) admits a power-law expansion around $x_0$. After that we prove that
for a large subclass of models  the speed of sound of the dark matter
component can be as low as required for structure formation.

We then move on and discuss an exactly solvable quadratic toy model, which is
undoubtedly of formal interest because it shows explicitly that the dust epoch
is not necessarily associated with an extremum in the pressure.

Finally, we revisit an exact transient model (a sort of modified Chaplygin
gas), which  describes successfully universes dominated by clustering dark
matter at early times.

\section*{Acknowledgments}

L.P.C. is partially funded by the University of Buenos Aires  under
project X223, and the Consejo Nacional de Investigaciones Cient\'{\i}ficas y
T\'ecnicas. R.L. is  financially supported by the University of the Basque Country through research grant 
UPV00172.310-14456/2002, by the Basque Government through fellowship BFI01.412, and by the former Spanish Ministry of Science and Technology
 through research grant  BFM2001-0988. 

\end{document}